\documentclass[10pt,preprint]{emulateapj}

\shorttitle{An X-ray Variable Millisecond Pulsar}
\shortauthors{Bogdanov, Grindlay, \& van den Berg}

\begin{document}

\title{An X-ray Variable Millisecond Pulsar in the Globular Cluster 47
Tucanae: \\ Closing the Link to Low Mass X-ray Binaries}

\author{Slavko Bogdanov, Jonathan E. Grindlay, Maureen van den Berg}

\affil{Harvard-Smithsonian Center for Astrophysics, 60 Garden Street,
Cambridge, MA 02138 \\ sbogdanov@cfa.harvard.edu,
josh@cfa.harvard.edu, mvandenberg@cfa.harvard.edu}

\begin{abstract}
We report the discovery of peculiar X-ray spectral variability in the
binary radio millisecond pulsar PSR J0024--7204W in the globular
cluster 47 Tucanae.  The observed emission consists of a dominant
non-thermal component, which is eclipsed for a portion of the orbit,
and a thermal component, which appears to be persistent.  We propose
that the non-thermal X-rays originate in a relativistic intrabinary
shock, formed due to interaction between the relativistic particle
wind from the pulsar and matter from the main-sequence companion star,
while the thermal photons are from the heated magnetic polar caps of
the millisecond pulsar. At optical wavelengths, the emission exhibits
large-amplitude variations at the orbital period, which can be
attributed to heating of one side of the tidally-locked secondary star
by the pulsar wind.  The observed X-ray and optical properties of PSR
J0024--7204W are remarkably similar to those of the low mass X-ray
binary and X-ray millisecond pulsar SAX J1808.4--3658 in
quiescence. This supports the conjecture that the non-thermal X-ray
emission and optical modulations seen in the SAX J1808.4--3658 system
in a quiescent state are due to interaction between the wind from a
reactivated rotation-powered pulsar and matter from the companion
star. The striking similarities between the two systems provide
support for the long-sought connection between millisecond radio
pulsars and accreting neutron star systems.
\end{abstract}


\section{INTRODUCTION}

Radio millisecond pulsars (MSPs) form a separate class of
rotation-powered pulsars, characterized by small spin periods,
$P\lesssim25$ ms, and spin-down rates, $\dot{P}=dP/dt\sim10^{-19-21}$,
implying surface magnetic field strengths of
$B\propto(P\dot{P})^{1/2}\sim10^{8-9}$ Gauss and characteristic ages of
$\tau\equiv P/2\dot{P} \gtrsim 10^9$ years. These neutron stars (NSs)
are widely believed to be the end products of a more exotic channel of
binary evolution, involving an extended period of accretion of matter
and angular momentum from a close stellar companion in X-ray binary
systems \citep{Alp82,Bhatt91}. After a period of $10^{7-8}$ years,
they acquire millisecond spin periods and after accretion ceases are
reactivated as radio pulsars. This theory is supported by the fact
that over $\sim$75\% of the known MSPs in the Galactic disk are found
in binary systems, whereas this is the case for only $\sim$~1\% of the
general pulsar population\footnote{Australia Telescope National
Facility Pulsar Catalog, available at
http://www.atnf.csir.au/research/pulsar/psrcat}. The strongest support
so far for the connection between MSPs and X-ray binaries comes from
the discovery of coherent millisecond X-ray pulsations in the low mass
X-ray binary (LMXB) transient SAX J1808.4--3658 \citep{Wij98} and
subsequently five\footnote{For the discovery of the latest X-ray
millisecond pulsar see http://www.astronomerstelegram.org/?read=353}
other systems \citep[see][for a review]{Camp98}.  However, a
firm connection between the two classes of objects is yet to be
established as no LMXB has been detected as a radio pulsar and no MSP
has shown X-ray properties of a quiescent LMXB (qLMXB).

The globular cluster 47 Tucanae (NGC 104, henceforth 47 Tuc) is known
to host at least 22 radio millisecond pulsars \citep[and references
therein]{Camilo00,Freire01,Freire03,Camilo05}.  Many of these MSPs,
however, have flux densities below the sensitivity of the Parkes radio
telescope used to detect them and are only observable on rare
occasions when strong interstellar scintillation focusing effects
bring them above the detection limit.  Thus, to date, 17 of the 22
MSPs have been detected sufficiently frequently to allow a precise
determination of their positions from radio timing solutions. During
the only radio observation in which PSR J0024--7204W (henceforth 47
Tuc W or MSP W) was detected\footnote{After submission of this article
for review, it was brought to our attention that this MSP has been
detected again (P. Freire, private communication).}, this
2.35-millisecond pulsar was found to be in a 3.2 hour binary orbit and
to undergo eclipses but its position could not be determined
accurately \citep{Camilo00}.  Fortunately, \textit{Hubble Space
Telescope} (\textit{HST}) observations of 47 Tuc have been used to
match the binary period and phase of the modulations of a variable
star with that of MSP W and thus determine the position of the MSP
with sub-arcsecond precision \citep{Edm02}. Surprisingly, the optical
counterpart of 47 Tuc W was found to be a $\gtrsim$$0.13$ M$_{\odot}$
main sequence star, which is unexpected for a MSP binary companion
since in all but one other case the secondary star is greatly evolved.
This intriguing result implies that either the companion is the star
that very recently finished spinning up MSP W or that it is not the
original secondary star but was instead exchanged for the low-mass
($<$0.1 M$_{\odot}$) companion, which, in the process, was liberated
from the binary.  The \textit{HST} identification showed MSP W to be
positionally coincident with the X-ray source W29, first detected by a
72-kilosecond \textit{Chandra X-ray Observatory} ACIS-I observation of
the core of 47 Tuc \citep{Grind01}. The X-ray spectrum was found to
be relatively hard, suggestive of non-thermal emission, atypical of
the MSP sample in 47 Tuc, but similar to PSR J1740--5340 in the
globular cluster NGC 6397 \citep{DAm01,Grind02}, the only other known
MSP with a non-degenerate stellar companion.

\section{OBSERVATIONS AND DATA ANALYSIS}

We have employed the superb spatial resolution of \textit{Chandra} and
\textit{HST} to further investigate the unusual properties of the 47
Tuc W system. The X-ray dataset consists of four additional
65-kilosecond observations of the core of 47 Tuc performed with the
\textit{Chandra} ACIS-S instrument. These deep observations have
resulted in the detection of $\sim$300 sources within the half-mass
radius of 47 Tuc \citep{Hein05} including the X-ray counterparts for
each of the MSPs with known positions \citep{Bog05}. The optical
\textit{HST} data consist of three sets of archival observations
acquired with the High Resolution Camera (HRC) on the Advanced Camera
for Surveys (ACS) and one set of observations performed
contemporaneously with the \textit{Chandra} observations using the
Wide Field Camera (WFC) on the ACS.  Table 1 summarizes all X-ray and
optical observations used in this analysis.

\begin{table}
\begin{center}
\tabletypesize{\small}
\tablecolumns{3} 
\tablewidth{0pc}
\caption{A summary of the X-ray (\textit{Chandra}) and optical (\textit{HST}) observations used in this analysis.}  
\begin{tabular}{lcc}
\hline
\hline
Telescope/ & Epoch of & Observation \\
Instrument & Observation & ID \\ 
\hline

Chandra/ACIS-S & 2002 Sep 29 & 2735 \\
Chandra/ACIS-S & 2002 Sep 30 & 2736 \\
Chandra/ACIS-S & 2002 Oct ~2 & 2737 \\
Chandra/ACIS-S & 2002 Oct 11 & 2738 \\
HST/ACS-HRC & 2002 Apr ~5  & 9028\tablenotemark{a} \\
HST/ACS-HRC & 2002 Apr 12-14 & 9019 \\
HST/ACS-HRC & 2002 Jul 24 & 9443 \\
HST/ACS-WFC & 2002 Sep 30$-$Oct 11\tablenotemark{b} & 9281 \\ 

\hline
\end{tabular}
\tablenotetext{a}{See also Edmonds et al. (2002).}
\tablenotetext{b}{Simultaneous with the \textit{Chandra} ACIS-S
observations.}
\end{center}
\end{table}

The initial data reduction and image processing of the X-ray
observations were performed using the CIAO\footnote{Chandra
Interactive Analysis of Observations, available at
http://asc.harvard.edu/ciao.} 3.0 software package and are discussed
in detail elsewhere \citep{Hein05}. For the spectral and variability
analyses, the \textit{Chandra} observations were first filtered to
only include events with energies between 0.3 and 8.0 keV.
Subsequently, the X-ray photons from MSP W were obtained from a
circular region with a $1''$ radius around the position of the optical
counterpart as this circle encloses $>$90\% of the total energy for an
assumed 0.5-1.0 keV thermal source. The net count rate was then
obtained by subtracting a background taken from three source-free
regions on the image. To facilitate the spectral analysis, the
$\sim$300 net photons were grouped into energy bins, while requiring
that each bin contain at least 15 counts. For all spectral fits the
hydrogen column density toward 47 Tuc was fixed at $N_{\rm H}=(1.3 \pm
0.3)\times10^{20}$ cm$^{-2}$ \citep{Gratt03,Card89,Pred95}. An X-ray
lightcurve was obtained by folding the observations at the binary
period using the latest ephemeris obtained from radio pulse timing
observations (Freire et al., unpublished data).

The optical colors and variability of 47 Tuc W were first investigated
in our ACS/WFC data. Unfortunately, due to the large pixelscale of the
WFC ($0\farcs{049}$ per pixel) and the presence of several relatively
bright stars around the position of 47 Tuc W, the faint optical source
was not detected in the individual ACS observations. Since the images
were dithered relative to each other with sub-pixel offsets, we could
create combined images in each filter with improved spatial resolution
using the Space Telescope Science Institute multidrizzle routines. In
these phase-averaged images, the companion of 47 Tuc W is visible near
the limit of detection. However, the large errors in the magnitude
($\sim$0.3--0.5 mag) prevented us from obtaining reliable lightcurves
or a significant limit on the H$\alpha$ emission expected from this
system. Thus, we have used the higher resolution ($0\farcs{027}$ per
pixel) ACS/HRC archival data, where 47 Tuc W is clearly visible when
at maximum brightness.  We measured magnitudes using aperture
photometry on the distortion-corrected pipeline-processed images. A
small aperture (2 pixel radius) was chosen to minimize contamination
by the flux of nearby stars. For a few frames, mostly for those in
which the counterpart of 47 Tuc W was near minimum brightness, a small
change in the choice of background regions resulted in magnitude
differences of up to 0.8 mag. We corrected these magnitudes to an
aperture of 6 pixels using aperture photometry of several bright and
relatively isolated stars within 200 pixels from the optical
counterpart of 47 Tuc W. For each filter, we checked the variation of
this correction as a function of time but found no indication for a
systematic trend. This led us to apply a time-averaged aperture
correction per filter. Next, we used the filter-dependent encircled
energy curves from \citet{Sir05}\footnote{Available at
http://acs.pha.jhu.edu/instrument/calibration/photometry/.}, that give
the fraction of total source counts as a function of aperture radius,
to correct the magnitudes to an infinite aperture \citep[$5\farcs5$,
see][for details]{Sir05}. We consider as unreliable those measurements
for which these differences would result in flux differences of more
than $\sim$30\% and removed them from our dataset. The remaining data
points do not show a systematic difference when compared to magnitudes
derived for a different choice of background.  Instrumental magnitudes
were converted to magnitudes in the STmag system using the most recent
filter-dependent
sensitivities\footnote{http://www.stsci.edu/hst/acs/analysis/zeropoints.}.
We used these fluxes to obtain sinusoidal fits to the lightcurves and
make phase-dependent spectral energy distributions (SEDs). The times
of mid-observation were converted to photometric phases using the
existing radio ephemeris (Freire et al., unpublished data).
 
\clearpage

\section{THE X-RAY SPECTRUM} 

The X-ray spectrum of 47 Tuc W in the 0.3--8.0 keV range suggests that
the emission consists of at least two components: a dominant
non-thermal (power-law) and a fainter thermal component. The latter is
fit equally well by either a simple blackbody (BB) or a NS hydrogen
atmosphere (NSA) model \citep{Rom87,Zavlin96,Lloyd03}.  After
excluding the photons between orbital phases $\phi=0.4-0.7$ to correct
for the bias in the best fit parameters introduced by the variability
of the non-thermal component (see next section and Fig. 1), we obtain
the following best fit models.  If we assume a BB model for the
thermal component, we find that the spectrum can be described by a
non-thermal component with a power law photon index of $\Gamma
=1.13\pm0.35$ and a BB with an effective temperature of $T_{\rm
eff}=(1.56\pm0.28) \times 10^6$ Kelvin and an effective emitting
radius of $R_{\rm eff}=0.45\pm0.43$ km ($\chi_{\nu}^2=1.22$ for 14
degrees of freedom), with a total unabsorbed flux of
$F_{X}=(1.4\pm0.3)\times10^{-14}$ ergs cm$^{-2}$ s$^{-1}$. On the
other hand, if we consider a NSA model and assume a gravitational
redshift of $z_{g} = 0.31$ at the NS surface, we obtain
$\Gamma=1.14\pm0.35$, $T_{\rm eff}=0.95^{+0.31}_{-0.24}\times10^6$ K,
$R_{\rm eff} = 0-2.3$ km ($\chi^2_{\nu}=1.17$ for 14 degrees of
freedom), and a total unabsorbed flux of $F_{X} = (1.5\pm0.3)\times
10^{-14}$ ergs cm$^{-2}$ s$^{-1}$.  In both cases, the non-thermal
component contributes with $\sim$75\% of the total flux. For a
distance to 47 Tuc of 4.85 kpc \citep{Gratt03} we obtain a total X-ray
luminosity of $L_{X}=(4\pm1)\times10^{31}$ ergs s$^{-1}$ and a
non-thermal luminosity of $L_{\rm X,NT}\approx(2-3)\times10^{31}$ ergs
s$^{-1}$ (0.3--8.0 keV). All errors quoted above represent $1\sigma$
uncertainties in the derived parameters. We note that although an
F-test does not indicate that a composite model is statistically
preferred over a pure power-law model (but with a steeper spectral
index of $\Gamma=1.7\pm0.1$), the X-ray lightcurve, discussed in \S 4,
suggests that the thermal component in the spectrum is genuine.  In
addition, the best fit values for $T_{\rm eff}$ and $R_{\rm eff}$ are
quite similar to what is observed in MSPs with predominantly thermal
spectra, such as the majority of MSPs in 47 Tuc \citep{Grind02,Bog05}.
The marginally acceptable fits suggest that a second thermal component
is probably present in the spectrum, as seen in the nearby MSPs
J0437--4715 and J0030+0451 \citep{Zavlin02,Beck02}.

Thermal emission from MSPs is believed to originate from the magnetic
polar caps of the underlying NS, heated by a backflow of energetic
particles from the pulsar magnetosphere \citep[see e.g.][and
references therein]{Hard02a,Zhang03}. On the other hand, non-thermal
emission from MSPs can be produced by two different mechanisms. In the
most luminous MSPs, B1821--24 and B1937+21 \citep{Beck99}, the narrow
X-ray pulse profiles indicate that the non-thermal emission is beamed
and must originate in the pulsar magnetosphere, near the light
cylinder, and is probably synchrotron and curvature radiation from
charged particles accelerated to relativistic speeds in the strong
magnetic field. Alternatively, non-thermal radiation can be produced
by the interaction of the pulsar wind with the binary companion or the
interstellar medium.  The resulting shock wave is expected to be a
prominent source of high-energy photons, via the synchrotron or
inverse Compton radiation produced by the highly accelerated particles
at the shock front and in the flow downstream from the shock
\citep{Arons93}.

\section{X-RAY VARIABILITY}

 In the 47 Tuc W system, we do not know \textit{a priori} the exact
nature of the observed non-thermal emission. Although the Chandra
ACIS-S observations preclude pulsation analysis due to the limited
time resolution (3.2 s), the longer-timescale variability contains a
wealth of information about the origin of the non-thermal X-rays. In
particular, the lightcurve folded at the binary period exhibits
dramatic variations as a function of orbital position (upper panel of
Fig. 1).  Based on Poisson statistics, we find that the observed
variability is significant at the 99.9\% confidence level.

\begin{figure}
\figurenum{1}
\epsscale{1.0}
\includegraphics[width=0.48\textwidth]{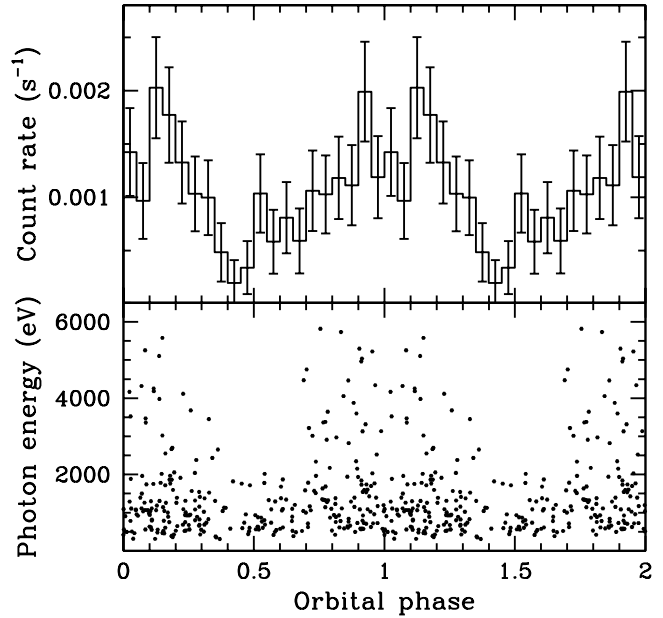}
\caption{{\it (Upper panel)} X-ray lightcurve of PSR
J0024--7204W in the 0.3-6.0 keV band, folded over its binary
period. Note the X-ray minimum at $\phi\approx0.45$ instead of
$\phi=0.5$, the position of the optical minimum. The uncertainty in
the absolute phase is $>$$10^{-4}$. {\it (Lower panel)} The energy of
each photon detected within $1''$ of the optical position of 47 Tuc W
versus orbital phase.  The data shown in this plot cover an integer
number of binary orbits (21) to ensure uniform phase coverage. Note the
remarkable absence of photons with energies above 2 keV and the
reduction in the number of soft photons between orbital phases 0.4 and
0.7. Two orbital cycles are shown for clarity.}
\end{figure}

Surprisingly, a comparison of the X-ray colors from different orbital
phases shows that the spectrum of MSP W softens significantly around
the minimum for $\sim$30\% of the orbit, with no counts detected above
2 keV (see lower panel of Fig. 1). In addition, at phases between
0.4 and 0.7, we find a decline of about 70\% in the number of
soft photons ($<$2 keV).  The remaining $\sim$30\% are consistent with
being entirely due to the underlying thermal (BB or NSA) component
although some of these photon may be due to partial eclipses of the
non-thermal source. In any case, there is strong indication that
the non-thermal component is occulted for roughly one third of the
orbit, most likely by the main-sequence secondary star. Note that the
underlying X-ray source cannot be eclipsed by diffuse material within
or around the binary, as this would be manifested primarily by
absorption of soft photons.
 
The characteristics of the X-ray spectrum and lightcurve are
suggestive of emission originating from two different locations within
the binary.  As stated above, the thermal component most likely
originates from the heated polar caps of the MSP.  On the other hand,
the observed X-ray eclipses are inconsistent with a magnetospheric
origin of the non-thermal photons. If these X-rays did originate from
the pulsar magnetosphere, one would expect both emission components to
undergo much shorter (at most 2-5\% of the orbit) total eclipses,
characterized by very rapid ingress and egress and centered at exactly
$\phi=0.5$. The observed eclipses also suggest a localized non-thermal
source, rather than an extended circum-binary X-ray emission region
that could form, for instance, by interaction of the pulsar wind with
the interstellar medium.

The behavior of the hard X-ray emission can be naturally explained by
the presence of a relativistic shock wave within the binary, produced
by interaction between the pulsar wind and matter from the secondary
star. This shock forms at some distance from the MSP where the ram
pressure of the pulsar wind balances the pressure exerted by the gas
from the companion star. The length of the X-ray eclipse ($\sim$30\%
of the orbit) suggests that this occurs much closer to the companion
star than to the MSP. The shock can then be eclipsed by the secondary
for a substantial portion of the orbit, even if the MSP is never
occulted. As noted above, if the MSP is occulted by the secondary,
which would occur for $80^{\circ}\lesssim i \le 90^{\circ}$, the
eclipse should last for no more than 2-5\% of the
orbit. Unfortunately, the limited number of photons observed at the
X-ray minimum (only 3 detected counts in a bin of width 10 minutes)
prevent us from determining whether this is indeed the case and thus
limiting the maximum allowed binary inclination.

If the secondary star is underfilling its Roche lobe (RL), the
observed shock could result from the pulsar wind impinging on and
evaporating material off of the surface of the companion or
interacting with the stellar wind, generated due to the co-rotation
forced by tidal coupling of the orbital motion to the deep
convection layers of the secondary star \citep[see e.g.][and
references therein]{Wij93}.  More likely, the main-sequence companion
may be overflowing its RL, as a result of bloating due to irradiation
by the energetic particle wind from the MSP \citep[see
e.g.][]{Pod91,DAn93,Vil94}.  The shock can then be produced by the
interaction of the pulsar wind with a stream of gas from the companion
issuing through the inner Lagrange point (L$_1$).  If this is indeed
the case, it permits us to place a conservative lower limit on the
orbital inclination by noting that at $\phi\approx0.45$, the
non-thermal X-ray source is completely occulted.  For the RL geometry
of a 1.4 M$_{\odot}$ MSP and a 0.13-0.29 M$_{\odot}$ companion,
corresponding to the allowed range of inclinations $90^{\circ}\ge i
\gtrsim 27^{\circ}$ \citep{Edm02}, we find $i\gtrsim35^{\circ}$.

An interesting feature of the X-ray eclipse is the apparent sharp
increase in the count rate immediately after the X-ray minimum. If
real, this emission may be the signature of Doppler boosting of the
particle flow downstream from the shock as it travels around the
secondary star \citep{Arons93}.  This effect can also account for the
possible drop in the countrate at $\phi\approx1$, corresponding to
the portion of the orbit at which the particles in the shock are
accelerated primarily in the direction away from the observer. Another
notable property of the eclipse is its asymmetry, characterized by a
rapid ingress and slower egress. In addition, the minimum in the X-ray
flux occurs at $\phi=0.4-0.45$, i.e. before the optical minimum at
superior conjunction ($\phi=0.5$). This suggests the presence of a
swept-back shocked region that is elongated perpendicular to and is
asymmetric about the semi-major axis of the binary. Such a geometry
arises because as the gas from the secondary interacts with the pulsar
wind it can no longer co-rotate along with the secondary star and
must, therefore, fall behind.  Figure 2 illustrates the likely
appearance of the non-thermal emission region for the RL overflow case
as inferred from the X-ray lightcurve.  The geometry of the shock is
similar if the companion star is contained within its RL except that
for the case of evaporative mass loss the shocked ``cometary'' tail
would originate at or very near the surface of the star.  We note that
a wind from the secondary star could also result in mass outflow
through L$_1$ and a configuration similar to that shown in Figure 2.
In all cases, when viewed at $\phi\approx0.4$, the apparent length of
the shocked region is relatively small, explaining the short duration
of ingress. Conversely, at $\phi\approx0.6$ the projected length of
the emission region is larger, causing a gradual transition out of
eclipse.  The relative durations of ingress and egress as well as the
length of the eclipse permit us to roughly constrain the geometry of
the gas stream and the shock.  Taking $i=90^{\circ}$, and a shock
approximately at L$_1$, we find an upper limit on the length of the
shocked stream of $l\lesssim2\times10^{10}$ cm, comparable to the
radius of the companion, and a distance of $d\lesssim5\times10^{9}$
cm, from the surface of the companion to the shock.

\begin{figure}
\figurenum{2}
\epsscale{1.0}
\includegraphics[width=0.48\textwidth]{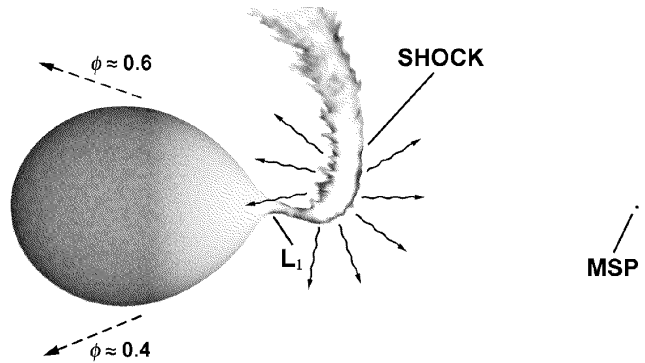}
\caption{A representation of the 47 Tuc W binary illustrating the key
features of the system for the case of a Roche lobe filling companion,
for assumed masses of 1.4 M$_{\odot}$ and 0.15 M$_{\odot}$ for the MSP
and main sequence companion, respectively. Here, the orbit is depicted
face-on ($i = 0^{\circ}$) to better show the geometry of the X-ray
emitting shocked region. The sense of rotation of the system is
counterclockwise and is in the plane of the page. The dashed arrows
point in the general direction of the observer roughly before
(\textit{bottom}) and after (\textit{top}) the X-ray minimum.}
\end{figure}

\clearpage

\section{LARGE-AMPLITUDE OPTICAL VARIABILITY}

\begin{figure}
\figurenum{3}
\epsscale{1.0}
\includegraphics[width=0.48\textwidth]{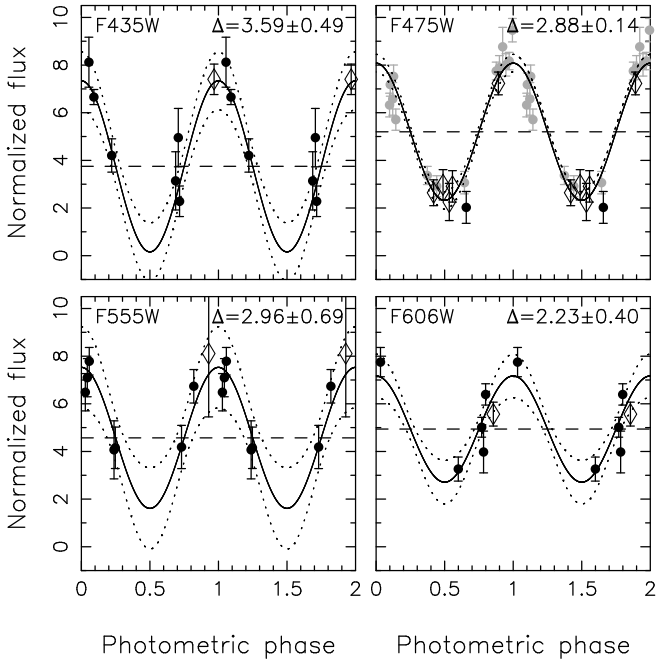}
\caption{\textit{HST} HRC light curves of the optical
counterpart of 47 Tuc W in different filters folded at the photometric
period. The flux is in units of 10$^{-18}$ ergs s$^{-1}$ cm$^{-2}$
\AA$^{-1}$. The data from observing programs 9019, 9028, and 9443 are
plotted with solid circles, gray circles, and open diamonds,
respectively. For each filter the solid line is the sine curve with
the best-fit amplitude and average flux level, with period and phasing
from the most recent radio timing ephemeris (Freire et al.,
unpublished data). The dotted lines represent 2$\sigma$ error margins
while the dashed lines give the average flux levels. The \textit{HST}
filter and $\Delta$, the amplitude of variation ($\pm$1$\sigma$), are
given in the upper left corner of each panel. Note the trend of
decreasing amplitude with increasing wavelength. Two orbital cycles
are shown for clarity.}
\end{figure}

Figure 3 shows the \textit{HST} ACS/HRC light curves in the F435W,
F475W, F555W, and F606W bands. For these filters there is sufficient
phase coverage to fit the amplitude and average flux level of
variation. For the fit we assumed a sinusoidal light curve and the
latest radio ephemeris (Freire et al., unpublished data). To create
SEDs, the fitted fluxes at maximum, mean, and minimum brightness were
first corrected for the extinction towards 47 Tuc using the
filter-dependent conversions in Table 32 of \citet{Sir05} for the
spectral type (G2) closest to that of the counterpart of MSP W (see
below for the results of our blackbody fits). Since the detailed
spectrum of the counterpart is unknown, it is difficult to estimate
the corrresponding systematic errors. From comparison with the only
two conversions provided for other spectral types (O5 and M0), we
expect that these are small. The effective wavelengths that are used
in the SEDs are derived with the task {\tt calcphot} in
IRAF/STSDAS\footnote{IRAF is distributed by the National Optical
Astronomy Observatories, which are operated by the Association of
Universities for Research in Astronomy, Inc., under cooperative
agreement with the National Science Foundation.}  from folding the
spectrum of a reddened [$E(B-V)=0.023$] $5500$ K blackbody with the
filter response curves.  The resulting SEDs for photometric phases
$0$, $0.25/0.75$, and $0.5$ were then fit to a BB spectrum with fixed
$R_{\rm eff}$ but variable $T_{\rm eff}$.

We find that for the SED at maximum brightness, the data are best fit
with $T_{\rm eff}$ between $\sim$$6800$ K, for $R_{\rm eff}=0.15$
R$_{\odot}$ (the radius appropriate for an unheated 0.13 M$_{\odot}$
star) and $\sim$$5400$ K for $R_{\rm eff}=0.26$ R$_{\odot}$
(corresponding to the RL radius for $i=90^{\circ}$ and $M_{\rm
MSP}=1.4$ M$_{\odot}$). Similarly, the fit to the SED at mean
brightness results in a range of $T_{\rm eff}\sim6200-5000$ K, while
at minimum brightness we obtain $T_{\rm eff} \sim 5200-4400$ K for the
same range of stellar radii. We note that although the fits are
consistent with $R_{\rm eff}< R_{\rm RL}$, this does not necessarily
mean that the companion is not RL filling.  It is more likely that
only a portion of the hemisphere facing the MSP is actually heated due
to less efficient heating near the poles (see Fig. 2). Furthermore,
small apparent values for R$_{\rm eff}$ can result from a viewing
angle effect for $i<90^{\circ}$.  Given the uncertainties in the
inclination and the details of particle heating, we can only constrain
the temperature difference between the heated and unheated areas of
the secondary to $\Delta T_{c}\sim1000-2400$ K, under the assumption
that $R_{\rm eff}$ at optical maximum is less than or equal to $R_{\rm
eff}$ at minimum.

\begin{figure}
\figurenum{4}
\includegraphics[angle=270,width=0.48\textwidth]{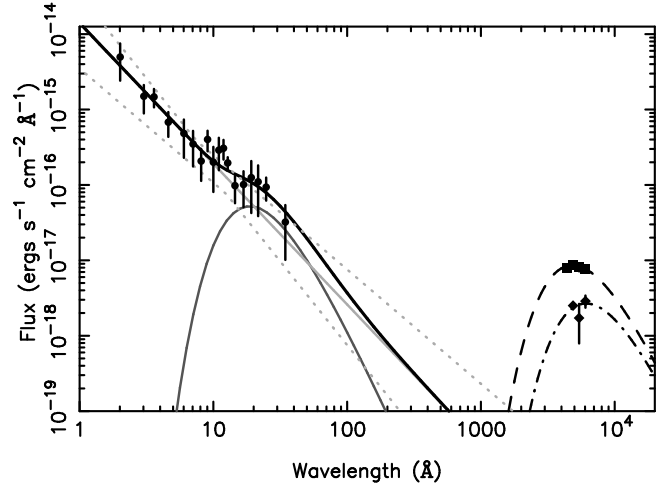}
\caption{The broadband spectrum of 47 Tuc W showing the X-ray data
(\textit{circles}) and optical data at maximum (\textit{squares}) and
minimum (\textit{diamonds}). Also shown is the best fit model for the
total emission (\textit{heavy black line}), as well as for the
individual emission components: thermal emission from the MSP
(\textit{dark grey line}), blackbody emission from the secondary star
at optical maximum and minimum (\textit{dashed} and \textit{dot-dashed
line}, respectively) assuming $R_{\rm eff}=0.20$ R$_{\odot}$, and
non-thermal emission from the intrabinary shock (\textit{light grey
line}). The dotted lines delineate the bounds of the 1$\sigma$
uncertainty in the best fit non-thermal (power-law) model.}
\end{figure}

An extrapolation of the non-thermal X-ray emission to optical
wavelengths shows that the non-thermal emission produced in the
intrabinary shock contributes negligibly to the optical emission, even
if we take into account the large uncertainties in the absolute
magnitudes of the optical data and the non-thermal X-ray photon index
(see Fig. 4).  Therefore, the observed large-amplitude variability
must be due to a temperature difference of the two hemispheres of the
companion, which can be attributed to heating of one side of the
tidally-locked secondary star by the MSP, and not from the shock. This
scenario is consistent with the lightcurves in Figure 3, which show a
trend of decreasing amplitude of the optical modulation with
increasing wavelength. As we discuss later, both the overall X-ray to
optical spectrum and variability are very similar to those observed in
the SAX J1808.4--3658 system \citep{Camp02,Camp04}.

The combined X-ray and optical data permit us to determine the origin
of the energy supplied to the secondary star.  In particular, we find
that the total X-ray and UV flux from the intrabinary shock is
insufficient to heat the face of the companion star to the observed
temperature ($T_{H}\sim5400-6800$ K).  Therefore, we conclude that the
heating of the secondary must instead be due to the relativistic
particle wind of the MSP, generated at the expense of the rotational
energy of the underlying NS.  The rate of this energy loss is given by
$\dot{E}=4\pi^2I\dot{P}/P^3$, where $I$ is the NS moment of inertia,
typically assumed to be $10^{45}$ g cm$^2$.  Unfortunately, 47 Tuc W
is rarely detected at radio frequencies which does not allow a
determination of the pulsar spindown rate $\dot{P}$. Hence, we have no
direct measure of $\dot{E}$.  Nonetheless, we can use the correlation
between the observed thermal X-ray luminosity ($L_{\rm X,T}$) and
$\dot{E}$ found for the 47 Tuc MSPs, to obtain a crude estimate of
this parameter. Using $L_{\rm X,T}\sim(0.5-1)\times10^{31}$ ergs
s$^{-1}$ and the empirical relation $\log L_{\rm
X,T}=0.59\log\dot{E}+10.0$ \citep{Grind02}, we obtain
$\dot{E}\sim1\times 10^{35}$ ergs s$^{-1}$.  Assuming an isotropic
pulsar wind, we find that the total incident power on the secondary
star is $L_{\rm irr}\sim1\times10^{33}$ ergs s$^{-1}$. This value is
sufficient to heat the face of the companion star to the observed
luminosity ($L_{H}\sim1\times10^{32}$ ergs s$^{-1}$), if we allow for
a re-radiation efficiency of order 0.1. Furthermore, the flux from the
MSP wind at the surface of the secondary ($\sim$$10^{12}$ ergs
cm$^{-2}$ s$^{-1}$) is conducive to bloating of the star to a
significantly larger radius \citep[see Fig. 1 in][]{Pod91}, provided
that the particles constituting the wind deposit their energy below
the stellar photosphere. Given that the pulsar wind is probably
composed of highly relativistic electrons, positrons, and ions, which
are able to penetrate deep into the stellar interior, this condition
for expansion of the star can be satisfied, implying that the
secondary star is most likely RL filling.  Finally, the flux from the
pulsar wind at $L_1$ is consistent with the observed value of $L_{\rm
X,NT}\approx3\times10^{31}$ ergs s$^{-1}$, if we take into account the
particle acceleration efficiency $\epsilon_{a}\simeq0.2$
\citep{Arons93}, as well as the fraction of the wind interacting with
matter from the companion ($\sim10^{-3}$).

\section{EMISSION PROPERTIES OF THE INTRABINARY SHOCK}

The compact nature of the 47 Tuc W binary ensures that the intrabinary
shock is located in a relatively strong pulsar magnetic
field. Therefore, we expect synchrotron emission to be the principal
energy loss mechanism in the shock. However, the proximity of the
shock wave to the secondary star implies that the relativistic
electrons in the shock are immersed in a ``sea'' of optical photons
emanating from the secondary star. Thus, inverse Compton scattering
(ICS) may also be an important production mechanism for high-energy
photons.

The magnetic field strength immediately upstream of the shock is given
by $B_1=[\sigma/(1+\sigma)]^{1/2}(\dot{E}/cf_{p}d^2)^{1/2}$, where
$\sigma$ is the ratio of the magnetic energy flux to the kinetic
energy flux, $f_p$ is a geometric factor that defines the fraction of
the sky into which the pulsar wind is emitted, and $d$ is the distance
between the MSP and the shock \citep{Arons93}. For simplicity, we will
take $d\approx6.4\times10^{10}$ cm, corresponding approximately to the
distance from the MSP to L$_1$ assuming $M_{\rm MSP}=1.4$ M$_{\odot}$
and $i=60^{\circ}$.  For an isotropic pulsar wind ($f_{p}=1$), we
obtain $B_1\approx 2$ G and $B_1\approx 30$ G corresponding to the two
possible cases of a kinetic energy dominated ($\sigma=0.003$ as in the
case of the Crab nebula) and a magnetically dominated ($\sigma\gg 1$)
wind, respectively.  This implies a magnetic field strength beyond the
shock of $B_2=3B_1\sim 6$ G or $B_2\sim90$ G, respectively.

Production of the observed $\varepsilon=0.3-8$ keV photons via
synchrotron, therefore, requires relativistic electrons with Lorentz
factors
$\gamma=2.4\times10^5(\varepsilon/B_2)^{1/2}\sim1\times10^4-5\times10^5$.
From these values, we obtain a radiative loss time of $t_{\rm
synch}=5.1\times10^8(\gamma B_2^2)^{-1}\sim1-60$ s \citep{Ryb79}. If
we consider a power-law distribution of electron energies
i.e. $n_{e}(\gamma)\propto\gamma^{-p}$, where $p$ is related to the
photon index by $p=2\Gamma+1$ \citep{Ryb79}, and assume a roughly
cylindrical shocked emission region with length $l\sim2\times10^{10}$
cm and radius $\sim$$2\times10^{9}$ cm, we find that the required
electron density is $n_e\sim10^7$ or $n_e\sim10^4$ cm$^{-3}$,
corresponding to $\sigma=0.003$ and $\sigma\gg 1$, respectively.
These values imply energy densities in electrons of
$U_e\sim10^2-5\times10^3$ or $U_{e}\sim10^5-5\times10^6$ ergs
cm$^{-3}$.  The value for $\sigma\gg 1$ is comparable to the magnetic
energy density $U_{B}=B_2^2/8\pi\sim1-300$ ergs cm$^{-3}$.  Thus,
based on equipartition of energy arguments, the wind is most likely
magnetically dominated.

We note that if ICS were to dominate the X-ray emission, the radiating
electrons need to be only weakly accelerated to $\gamma\sim100$,
implying radiative loss times of $t_{\rm ICS}=(U_{B}/U_{\rm ph})t_{\rm
synch}\sim10-100$ hours, where $U_{\rm ph}=R_c \sigma_B
T_c^4/(R_{*}c)\sim 1$ ergs cm$^{-3}$ is the seed photon energy
density, while $R_{c}$, $T_{H}$, and $R_{*}$ are the radius,
temperature, and distance from the surface of the secondary star,
respectively.  The implied $U_{e}\sim10^7$ ergs cm$^{-3}$, however,
greatly exceeds the values of $U_{B}$ and $U_{\rm ph}$. Therefore, ICS
is an unlikely soft X-ray production mechanism in 47 Tuc W.

Future detailed multi-wavelength observations of the 47 Tuc W system
may reveal more information concerning the nature of the shock
emission, which, in turn, may provide insight into the little
understood properties of MSP winds, collisionless relativistic shocks,
and particle acceleration mechanisms. The existence of a regularly
eclipsed shock makes 47 Tuc W well suited for studies of these
phenomena.

\section{DISCUSSION}

\subsection{Comparison with Other MSPs}

The data presented here have revealed that 47 Tuc W exhibits X-ray
variability that is unique among the MSPs (as evident in
Fig. 1). These variations can be most easily explained by the presence
of a relativistic shock within the binary that is regularly eclipsed
by the secondary star. An intrabinary shock is also believed to be
present in PSR B1957+20, the so-called "black widow" eclipsing pulsar
\citep{Fru88}. This MSP has a non-thermal spectrum with $\Gamma =
1.9\pm0.5$ and $L_{X} = 2.7\times 10^{31}$ ergs s$^{-1}$ (0.5-7.0 keV)
\citep{Stap03}, for an assumed distance of 2.5 kpc \citep{Cord02}.
These values are comparable to those for MSP W and provide further
support for the presence of a shock in the 47 Tuc W system.  However,
we emphasize that 47 Tuc W is a fundamentally different binary than
PSR B1957+20 and similar eclipsing systems, such as 47 Tuc J, O, and R
(Bogdanov et al. 2005), which contain much less massive ($\sim$$0.03$
M$_\odot$), greatly evolved secondary stars.

MSP W posseses many characteristic similar to PSR J1740--5340 (J1740),
the only known MSP in the globular cluster NGC 6397
\citep{DAm01,Ferr03,Sabbi03}.  Specifically, J1740 exhibits radio
eclipses, X-ray emission which is seemingly hard and possibly variable
\citep{Grind02} and is bound to a $\sim$$0.3$ M$_\odot$ ``red
straggler'' companion that is currently filling its Roche lobe.  In
addition, the shape of the H$\alpha$ line observed in this 32.5-hour
binary system indicates the presence of a swept-back stream of gas
protruding from the companion, as the one possibly present in 47 Tuc W
(see Fig. 2).  However, unlike 47 Tuc W, J1740 shows no signature of
wind-driven heating of the companion in the photometric lightcurve
\citep{Kal03,Oro03}. This finding is consistent with the recently
revised value of $\dot{E}\approx3.3\times10^{34}$ ergs s$^{-1}$
\citep{Bassa04}, which indicates that the flux from the pulsar wind
incident on the $\sim$1.6 R$_{\odot}$ companion is too small to have a
measurable effect. Therefore, in this system RL overflow is not the
result of bloating by irradiation from the MSP but rather a
consequence of evolution off of the main sequence.  Perhaps the most
peculiar feature of the J1740 binary is the presence of He I
absorption localized in a thin longitudinal strip near the surface of
the secondary \citep{Ferr03}.  The shape of such a heated region could
be due to irradiation of the companion by a highly anisotropic pulsar
wind, preferentially emitted in the orbital plane. However, this would
require the MSP spin and orbital angular momentum vectors to be
exactly aligned.  The thin strip can be more plausibly explained by
the existence of an equatorial wind, emanating from the tidally-locked
companion star due to its forced rapid co-rotation, that is
interacting with the pulsar wind. Alternatively, the He I absorption
could originate in the swept back tail of material as the one shown in
Figure 2 for 47 Tuc W.

\subsection{The MSP-LMXB Connection}

The unusual X-ray and optical properties of 47 Tuc W can serve towards
finally establishing the long-suspected connection between LMXBs and
MSPs. This is now possible because the 47 Tuc W system appears to be
more typical of a qLMXB system, as it contains a main-sequence
companion that may be RL filling, than a MSP binary system. In
addition, the X-ray spectrum of 47 Tuc W system exhibits remarkable
similarities to that of the LMXB transient SAX J1808.4--3658
(henceforth J1808) in quiescence.  \textit{XMM-Newton} observations of
the latter system have revealed that during the long periods between
outbursts ($\sim$2 years), this 2.01-hour binary has an X-ray spectrum
which is seemingly purely non-thermal, with $\Gamma=1.4^{+0.6}_{-0.3}$
and an X-ray luminosity of $L_{X}=5\times10^{31}$ ergs s$^{-1}$ in the
0.5-10 keV band \citep{Camp02}.  Moreover, recent optical observations
of the bloated $\sim$0.05 M$_{\odot}$ brown dwarf companion of J1808
show sinusoidal variations at the orbital period \citep{Camp04}
implying a temperature difference of $\Delta T_{\rm c}
\sim$$1000\pm300$ K between the two faces of the companion, quite
similar to that reported here for 47 Tuc W.  Finally, the observed
non-thermal emission in J1808 also cannot account for the optical flux
as well as irradiating luminosity ($L_{\rm irr}\gtrsim4\times10^{33}$
ergs s$^{-1}$) required to produce the observed heating of the
secondary star \citep[see Fig. 1 of][]{Camp04}. The X-ray and optical
properties of 47 Tuc W and J1808 are compared in Table 2.

\begin{table}[t!]
\begin{center}
\tabletypesize{\small}
\tablecolumns{3} 
\tablewidth{0pc}
\caption{A comparison of the observed parameters for 47 Tuc W and
J1808.}
\begin{tabular}{lcc}
\hline
\hline
 & PSR J0024--7204W & SAX J1808.4--3658 \\ 
\hline
 $L_{X}$ (ergs s$^{-1}$) & $3\times10^{31}$ (0.3--8 keV) & $5\times10^{31}$ (0.5--10 keV) \\ 
 $\Gamma$ & $1.13\pm0.35$ (1$\sigma$) & $1.4^{+0.6}_{-0.3}$ (3$\sigma$) \\ 
 $\dot{E}$ (ergs s$^{-1}$) & $\sim$$10^{35}$ & ? \\ 
 $L_{i}$ (ergs s$^{-1}$) & $\sim$$1\times10^{31}$ & $\lesssim$$3\times10^{31}$ \\ 
 $L_{\rm irr}$ (ergs s$^{-1}$) & $\sim$$4\times10^{33}$ & $\gtrsim$$4\times10^{33}$ \\
 $T_{H}$ (K) & $5400-6800$ & $6200^{+850}_{-380}$ \tablenotemark{b} \\
 $L_{H}$ (ergs s$^{-1}$) & $\sim$$1\times10^{32}$ & $\sim$$4\times10^{31}$
 \\ $\Delta T_{c}$ (K) & $1000-2400$ & $\sim$$1000$ \\
\hline 
\end{tabular} 

\tablenotetext{a}{$L_{i}$ is the estimated intrinsic bolometric
luminosity of a 0.15 M$_{\odot}$ secondary star in the absence of
heating from the pulsar; $T_{H}$ and $L_{H}$ are the temperature
and bolometric luminosity of the heated face of the companion. The
other parameters listed are defined in the text.}
\tablenotetext{b}{S. Campana private communication.}

\end{center}
\end{table}

The primary difference between J1808 in quiescence and 47 Tuc W is the
lack of radio pulsations from the former. One possibility is that the
radio beams are not favorably oriented, rendering J1808 undetectable
at radio wavelengths. Alternatively, the radio signals from the
nascent MSP expected in J1808 could be obscured by a diffuse
circum-binary envelope of material, formed by the gas flowing out from
the companion that is ultimately expelled from the system by the
pulsar wind. Such an envelope is likely present around the 47 Tuc W
system as well and may explain, in part, why this MSP is occulted for
$\sim$40\% of the orbit and is rarely detected at radio frequencies
\citep{Freire05}. If this is the case, it would be yet another common
feature of the two systems.

We propose that the great similarities in the values of the power law
index and the non-thermal X-ray luminosity derived for 47 Tuc W and
J1808 point to a common origin of the non-thermal X-ray emission in
both systems. Also, it is highly probable that the source of energy
supplied to the heated companion in both systems is the same as
well\footnote{For J1808, unlike in the case of 47 Tuc W, no
simultaneous X-ray and optical data are available and the best limit
on $L_{X}$ is $<$$10^{34}$ ergs s$^{-1}$, as measured 2 days before
the optical observations.}.  If this is indeed the case, then, as
originally suggested by \citet{Bur03} and \citet{Camp04}, in the J1808
system there exists a pulsar wind, powered by the rotational energy of
the NS, which is constraining the outflow of material from the
companion and preventing its accretion onto the NS. As in the 47 Tuc W
system, the presence of this relativistic particle wind is manifested
via the non-thermal X-ray emission produced by interaction of this
wind with matter flowing out from the companion, as well as through
the optical variability caused by heating of one side of the
tidally-locked companion. By analogy, this would exclude an accretion
disk \citep[e.g][and references therein]{Gar01} or a propeller
mechanism as the source of hard X-rays in J1808 \citep[and references
therein]{Wij03}, since we know that in 47 Tuc W accretion is inhibited
by the pulsar wind.  The remarkable similarities between 47 Tuc W and
J1808 provide strong support for the claim that in episodes of
quiescence, the NS in the qLMXB system J1808 has already been
reactivated as a rotation-powered millisecond pulsar. This, in turn,
gives indirect evidence in favor of the long-standing hypothesis that
LMXBs are indeed progenitors of MSPs.

\subsection{X-ray Source Populations}

The implications of the 47 Tuc W system on studies of X-ray source
populations are profound since low luminosity ($L_{X}
\sim3\times10^{31}$ ergs s$^{-1}$) hard X-ray sources, which would
otherwise be classified as cataclysmic variables (CVs), i.e. accreting
white dwarf systems \citep{Grind01}, may, instead, be MSP binaries in
some cases.  The same holds true for the low luminosity, hard spectrum
qLMXBs categorized as black hole (BH) systems \citep{Gar01,Rut00}
based on the absence of thermal emission of a cooling NS.  The optical
heating effects of the pulsar wind, as seen in 47 Tuc W, would allow
MSPs to be discovered and distinguished from CVs or BH qLMXBs since
for both the low X-ray luminosities and the absence of an energetic
wind do not result in significant heating of the secondary star for
comparably short orbital periods. We note also that the discriminant
of X-ray to optical flux ratios, $F_{X}/F_{V}$, which might be used to
distinguish qLMXBs from CVs or MSPs is not useful for these transition
qLMXB-MSP systems: J1808 and MSP W have $\log(F_{X}/F_{V})\sim0.7$ and
$0.4$, respectively, which are typical values for CVs or qLMXBs but
much larger than a ``typical'' MSP with an evolved companion. For
comparison, the corresponding value for MSP U in 47 Tuc, whose
companion is likely a helium white dwarf \citep{Edm01}, is $\log
(F_{X}/F_{V})\sim-1.15$ \citep{Hein05}.\\


We thank C. O. Heinke for both his original analysis of the deep
\textit{Chandra} observations and many fruitful discussions, P.
Edmonds for discussions of his initial 47 Tuc W discovery analysis
using archival HST data, and P. Freire for providing the most
up-to-date binary ephemeris for MSP W.  We are also grateful to
F. Camilo and S. Campana for insightful discussions, D. Lloyd for use
of his neutron star atmosphere models, and R. Hynes, whose BinSim
software package was used to generate parts of Figure 2. This work was
supported in part by NASA \textit{Chandra} grant GO2-3059A and
\textit{HST} grant HST-GO-0944.


\begin{thebibliography}{}

\bibitem[Alpar et al.(1982)]{Alp82} Alpar, M. A., Cheng, A. F., M. A. Ruderman, M. A., \& Shaham, J. 1982, Nature, 300, 728

\bibitem[Arons \& Tavani(1993)]{Arons93} Arons, J. \& Tavani M. 1993, \apj, 403, 249

\bibitem[Bassa \& Stappers(2004)]{Bassa04} Bassa, C. G. \& Stappers, B. W. 2004, A\&A, 425, 1143

\bibitem[Becker \& Tr\"umper(1999)]{Beck99} Becker, W. \& Tr\"umper,
J. 1999, \apj, 341, 803

\bibitem[Becker \& Achenbach(2002)]{Beck02} Becker, W. \& Achenbach, B. 2002, Proceedings of the 270. WE-Heraeus Seminar on Neutron Stars, Pulsars, and Supernova Remnants, ed. W. Becker, H. Lech, \& J. Tr\"umper, p.64

\bibitem[Bhattacharya \& van den Heuvel(1991)]{Bhatt91} Bhattacharya, D. \& van den Heuvel, E. P. J. 1991, Phys. Rep. 203, 1

\bibitem[Bogdanov et al.(2005)]{Bog05} Bogdanov, S., Grindlay, J. E.,
Heinke, C. O., Camilo, F., \& Becker, W. 2005 in preparation.

\bibitem[Burderi et al.(2003)]{Bur03} Burderi, L., Di Salvo, T., D'Antona, F., Robba, N. R., \& Testa, V. 2003, A\&A, 404, L43

\bibitem[Camilo et al.(2000)]{Camilo00} Camilo, F., Lorimer, D. R., Freire, P., Lyne, A. G., \& Manchester, R. N.  2000, \apj, 535, 975

\bibitem[Camilo \& Rasio(2005)]{Camilo05} Camilo. F. \& Rasio, F. A. 2005, ASP Conf. Ser. Vol. 328: Binary Radio Pulsars, ed. F. A. Rasio \& I. H. Stairs (San Francisco: ASP), p. 147

\bibitem[Campana et al.(1998)]{Camp98} Campana, S., Colpi, M., Mereghetti, S., Stella, L., \& Tavani, M. 1998, A\&A Rev, 8, 279

\bibitem[Campana et al.(2002)]{Camp02} Campana, S., Stella, L., Gastaldello, F., Mereghetti, S., Colpi, M., Israel, G. L., Burderi, L., Di Salvo, T., \& Robba, R. N. 2002, \apj, 575, L15

\bibitem[Campana et al.(2004)]{Camp04} Campana, S., D'Avanzo. P., Casares, J., Covino, S., Israel, G. L., Marconi, G., Hynes, R., Charles, P., \& Stella, L. 2004, \apj, 614, L49

\bibitem[Cardelli, Clayton, \& Mathis(1989)]{Card89} Cardelli, J. A., Clayton, G. C., \& Mathis, J. S. 1989, \apj, 345, 245

\bibitem[Cordes \& Lazio(2002)]{Cord02} Cordes, J. M. \& Lazio, T. J. W. 2002, astro-ph/0207156 

\bibitem[D'Amico et al.(2001)]{DAm01} D'Amico, N., Lyne, A. G., Manchester, R. N., Possenti, A., \& Camilo, F. 2001, \apj, 548, L171

\bibitem[D'Antona \& Ergma(1993)]{DAn93} D'Antona, F. \& Ergma, E. 1993, A\&A, 269, 219

\bibitem[Edmonds et al.(2001)]{Edm01} Edmonds, P. D., Gilliland, R. L., Heinke, C. O. Grindlay, J. E., \& Camilo, F. 2001, \apj, 557, L57

\bibitem[Edmonds et al.(2002)]{Edm02} Edmonds, P. D., Gilliland, R. L., Camilo, F., Heinke, C. O., \& Grindlay, J. E. 2002, \apj, 579, 741

\bibitem[Ferraro et al.(2003)]{Ferr03} Ferraro, F. R., Sabbi, E., Gratton, R., Possenti, A., D'Amico, N., Bragaglia, A., \& Camilo, F. 2003, \apj, 584, 2003

\bibitem[Freire et al.(2001)]{Freire01} Freire, P. C., Camilo, F., Lorimer, D. R., Lyne, A. G., Manchester, R. N., \& D'Amico, N. 2001, \mnras, 326, 901

\bibitem[Freire et al.(2003)]{Freire03} Freire, P. C., Camilo, F., Kramer, M., Lorimer, D. R., Lyne, A. G., Manchester, R. N., \& D' Amico, N. 2003, \mnras, 340, 1359

\bibitem[Freire (2005)]{Freire05} Freire, P. C. C. 2005, ASP Conf. Ser. Vol. 328: Binary Radio Pulsars, ed. F. A. Rasio \& I. H. Stairs (San Francisco: ASP), p. 405

\bibitem[Fruchter, Stinebring, \& Taylor(1988)]{Fru88} Fruchter, A. S., Stinebring, D. R., \& Taylor, J. H. 1988, Nature, 333, 237

\bibitem[Garcia et al.(2001)]{Gar01} Garcia, M. R., McClintock, J. E., Narayan, R., Callanan, P., Barret, D., \& Murray, S. S. 2001, 553, L47

\bibitem[Gratton et al.(2003)]{Gratt03} Gratton, R. G., Bragaglia, A., Carretta, E., Clementini, G., Desidera, S., Grundahl, F., \& Lucatello, S. 2003, A\&A, 408, 529

\bibitem[Grindlay et al.(2001)]{Grind01} Grindlay, J. E., Heinke, C., Edmonds, P. D., \& Murray, S. S. 2001, Science, 292, 2290

\bibitem[Grindlay et al.(2002)]{Grind02} Grindlay, J. E., Camilo, F., Heinke, C. O., Edmonds, P. D., Cohn, H., \& Lugger, P. 2002, \apj, 581, 470

\bibitem[Harding \& Muslimov(2002)]{Hard02a} Harding, A. K. \& Muslimov, A. G. 2002, \apj, 568, 862

\bibitem[Heinke et al.(2005)]{Hein05} Heinke, C. O., Grindlay, J. E., Edmonds, P. D., Cohn, H. N., Lugger, P. M., Camilo, F., Bogdanov, S., \& Freire, P. C. 2005, \apj, in press (astro-ph/0503132)

\bibitem[Kaluzny, Rucinski, \& Thompson(2003)]{Kal03} Kaluzny, J., Rucinski, S. M., \& Thompson, I. B. 2003, \aj, 125, 1546

\bibitem[Lloyd(2003)]{Lloyd03} Lloyd, D. A. 2003, astro-ph/030356

\bibitem[Orosz \& van Kerkwijk(2003)]{Oro03} Orosz, J. A. \& van Kerkwijk, M. H. 2003, A\&A, 397, 237

\bibitem[Podsiadlowski(1991)]{Pod91} Podsiadlowski, Ph. 1991, Nature, 350, 136

\bibitem[Predehl \& Schmitt(1995)]{Pred95} Predehl, P. \& Schmitt, J. H. M. M. 1995, A\&A, 293, 889

\bibitem[Romani(1987)]{Rom87} Romani, R. W. 1987, \apj,313,718

\bibitem[Rutledge et al.(2000)]{Rut00} Rutledge, R. E., Bildsten, L., Brown, E. F., Pavlov, G. G., \& Zavlin, V. E. 2000, \apj, 529, 985

\bibitem[Rybicki \& Lightman(1979)]{Ryb79} Rybicki, G. B. \&  Lightman, A. P. 1979, Radiative Processes in Astrophysics (New York: Wiley)

\bibitem[Sabbi et al.(2003)]{Sabbi03} Sabbi, E., Gratton, R., Ferraro, F. R., Bragaglia, A., Possenti, A., D'Amico, N., \& Camilo, F. 2003, \apj, 589, L41

\bibitem[Sirianni et al.(2005)]{Sir05} Sirianni et al. 2005, PASP, submitted

\bibitem[Stappers et al.(2003)]{Stap03} Stappers, B. W., Gaensler, B. M., Kaspi, V. M., van der Klis, M., \& Lewin, W. H. G. 2003, Science, 299, 1372

\bibitem[Vilhu, Ergma, \& Fedorova(1994)]{Vil94} Vilhu, O., Ergma, E., \& Fedorova, A. 1994, A\&A, 291, 842

\bibitem[Wijers \& Paczy\'nski(1993)]{Wij93} Wijers, R. A. M. J. \& Paczy\'nski, B. 1993, \apj, 415, L115

\bibitem[Wijnands \& van der Klis(1998)]{Wij98} Wijnands, R. \& van der Klis, M. 1998, Nature, 394, 344

\bibitem[Wijnands(2003)]{Wij03} Wijnands, R. 2003, \apj, 588, 425

\bibitem[Zavlin, Pavlov, \& Shibanov(1996)]{Zavlin96} Zavlin, V. E., Pavlov, G. G., \& Shibanov, Yu. A., 1996, A\&A, 315, 141

\bibitem[Zavlin et al.(2002)]{Zavlin02} Zavlin, V. E., Pavlov, G. G., Sanwal, D., Manchester, R. N., Tr\"umper, J., Halpern, J. P., \& Becker, W. 2002, \apj, 569, 894

\bibitem[Zhang \& Chang(2003)]{Zhang03} Zhang, L. \& Chang, K. S. 2003, A\&A, 398, 639

\end{thebibliography}
\end{document}